\documentclass[]{spie}  

\usepackage{amsmath,amsfonts,amssymb}
\usepackage{graphicx}
\usepackage[colorlinks=true, allcolors=blue]{hyperref}
\usepackage{wrapfig}
\usepackage{lscape}
\usepackage{rotating}
\usepackage{epstopdf}
\usepackage{subcaption}

\title{Alignment of three mirror anastigmat telescopes using a multilayered stochastic parallel gradient descent algorithm}

\author[a,b]{Solvay Blomquist}
\author[a,b]{Heejoo Choi}
\author[a]{Hyukmo Kang}
\author[b]{Kevin Derby}
\author[b]{Pierre Nicolas}
\author[a]{Ewan S. Douglas}
\author[a,b]{Daewook Kim}
\affil[a]{Steward Observatory - The University of Arizona \linebreak 933 N Cherry Ave, Tucson, AZ 85721 \linebreak}
\affil[b]{Wyant College of Optical Sciences - The University of Arizona \linebreak 1630 E University Blvd, Tucson, AZ 85721}

\authorinfo{Further author information: (Send correspondence to H.C.)\\H.C. E-mail: hchoi@optics.arizona.edu}

\begin{document} 
\maketitle

\begin{abstract}
When a telescope doesn’t reach a reasonable point spread function on the detector or detectable wavefront quality after initial assembly, a coarse phase alignment on-sky is crucial. Before utilizing a closed loop adaptive optics system, the observatory needs a strategy to actively align the telescope sufficiently for fine wavefront sensing (WFS). This paper presents a method of early-stage alignment using a stochastic parallel-gradient-descent (SPGD) algorithm which performs random perturbations to the optics of a three mirror anastigmat telescope design. The SPGD algorithm will drive the telescope until the wavefront error is below the acceptable range of the fine adaptive optics system to hand the telescope over. The focused spot size over the field of view is adopted as a feed parameter to SPGD algorithm and wavefront peak-to-valley error values are monitored to directly compare our mechanical capabilities to our alignment goal of diffraction limited imaging and fine wavefront sensing.
\end{abstract}


\section{INTRODUCTION}
\label{sec:intro}  

In this study, we use an example three mirror anastigmat telescope (TMA), using a standard University of Arizona 6.5m f/1.25 mirror as used at the MMT observatory\cite{martin_fabrication_1997}, optimized to maximize image quality and  field of view in a constrained volume \cite{kim2023compact}. This TMA design is shown in Figure 1. Since TMA designs are notoriously more difficult to initially align than simpler two-mirror designs, we seek to prove the effectiveness of stochastic parallel gradient descent (SPGD) during the commissioning process of the telescope deployment. The goal of using SPGD in our case is to coarsely align the telescope, as the first stage of a sequence to reach better than or near diffraction-limited performance. We will require an alignment algorithm that can detect some random misalignment of the telescope system once it gets to its position where it will begin collecting science observations and then use the algorithm to adjust the available degrees of freedom and converge down to our defined threshold. 

In the context of SPGD, a quality metric is measured by adjusting a set of control parameters each iteration until a certain threshold is met \cite{vorontsov1998stochastic}. In our case, the measured quality metric each iteration will be the average root-mean-square (RMS) spot size radius measured across the field of view (FoV). The control parameters used to drive the convergence are the five degrees of freedom that we can adjust on M2.

To show the effectiveness of this algorithm when aligning the telescope during commissioning phase, we will present the methodology behind the algorithm. Following an explanation of the alignment, we will show the development in two phases, Phase I will show the correction algorithm with just the M2 hexapod misaligned and Phase II with both M1 and M2 misaligned. For simplicity in both phases, the M3-M4-focal plane is assumed to be pre-aligned and held in a fixed position while the five degrees of freedom that control the position of M2 can still be moved. 

\begin{figure}[hbt!]
    \centering
    \includegraphics[width=1.1\linewidth]{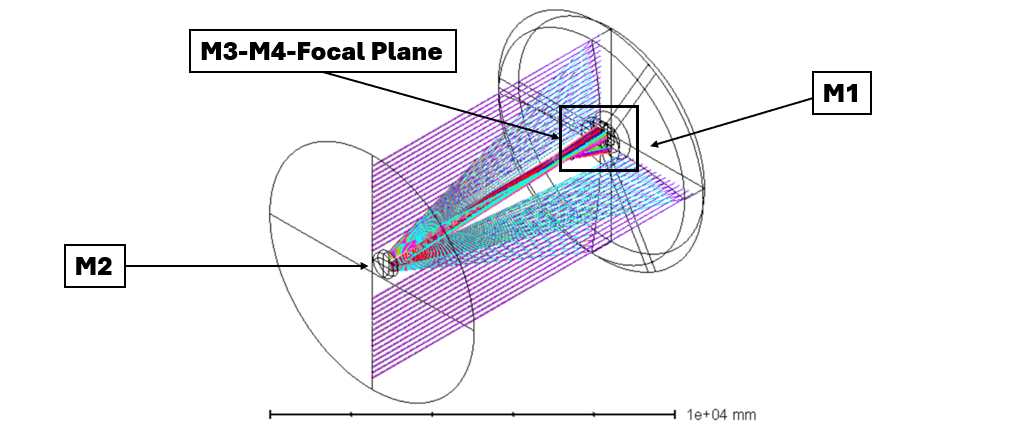}
    \label{fig:enter-label}
    \caption{Figure showing the optical design of the three mirror anastigmat (TMA) telescope system, reproduced from Kim et al. 2023. \cite{kim2023compact}}
\end{figure}

\section{Misalignment and Correction Algorithm}
\label{sec:sections}

When using SPGD during the commissioning phase of alignment, the measured quality metric evaluated at each iteration of the alignment loop is the average of the RMS spot size radius measured over our wide field of view. This is our current merit function evaluation for this algorithm. When applying random perturbations to the system for alignment on each iteration, the merit value will help us in understanding the required mechanical adjustment step and range for the alignments.  While evaluating the merit function at each iteration, the M2 position will be evaluated and updated based on improvement of the average spot size across FoV. When considering the wavefront error degeneracy between Decenter X and Tilt Y or Decenter Y and Tip X, we have separated the perturbations made when translating or tip/tilting M2 to be made on odd and even iterations, respectively. Figure 4 shows an overview of the flow of the process and the calculations made on each iteration. Previous studies have shown capabilities in correcting misalignment for a TMA telescope design caused by thermal perturbations to achieve diffraction limited performance. \cite{blomquist2023analysis}


\section{SPGD Phase I Results}

For Phase I of the SPGD algorithm, we are considering the alignment with no misalignments or corrections given to M1, we are only assuming misalignment for M2 and using only these parameters to align the system and converge to our defined requirement of $\sim2\lambda$ peak-to-valley (PV) wavefront error. As shown in Figure 2, with the misalignment given in Table 1, there is clear convergence from the initial misalignment and by iteration 500 of the algorithm reaches the defined requirement for a finer alignment to take control for science observations.


The two merit values being measured while applying random perturbations to the M2 hexapod will help us in understanding the required mechanical adjustment range for our random alignments during the commissioning phase of the telescope deployment. However, at this stage of the alignment on orbit, we cannot access a wavefront PV measurement. While developing Phase I, we found the estimated spot size radius corresponding to 2$\lambda$ PV WFE across the field is roughly 0.05 mm. More importantly, our alignment algorithms post SPGD have been able to successful continue alignment given a starting position where average RMS spot size radius is 0.05 mm. 

The goal of SPGD alignment at this stage of its development is to be able to show that we can meet our 2$\lambda$ PV WFE science requirement given some misalignment to M2 and using random perturbations of its position with infinite resolution on M2 motion. In Figure 2, we meet the goal of this phase by converging to this value within 500 iterations of the algorithm. Further optimization of SPGD, to be explored in Phase II, will include looking at field-dependent spot size trends and recovering the spot diagram symmetry and size convergence. The scheduled motion in M2 is implemented sequentially, and corresponding spot diagrams are recorded at the edge of the FoV and center. The polynomial fitting algorithm is used to find the optimal position of M2 corresponding to the small and symmetric spot diagram. The detailed steps are elaborated in the next section. In the iteration counting, this process is shown as single iteration but this one iteration will include multiple M2 motion and spot diagram evaluation steps. We have been able to pass the results from Phase I SPGD onto a finer but narrow measurable range phase retrieval alignment and further converge down to the requirement needed to perform science observations.

\begin{table}[hbt!]
    \centering
    \setlength{\tabcolsep}{10pt} 
    \renewcommand{\arraystretch}{1.5} 
    \begin{tabular}{|c|c|} \hline 
         \textbf{Degree of Freedom (DoF)}& \textbf{Misalignment (units)}\\ \hline 
         Decenter X & 3 (mm) \\ \hline 
         Decenter Y & 3 (mm) \\ \hline 
         Separation between M1 and M2 & 3 (mm) \\ \hline 
         Tilt X & 0.05 (deg.) \\ \hline 
         Tilt Y & -0.05 (deg.) \\ \hline
    \end{tabular}
    \caption{Initial misalignment values applied to the M2 before the SPGD optimization algorithm. This misalignment corresponds to the initial misalignment of the convergence plot shown in Figure 2.}
    \label{tab:my_label}
\end{table}

\begin{figure}[hbt!]
    \centering
    \includegraphics[width=1\linewidth]{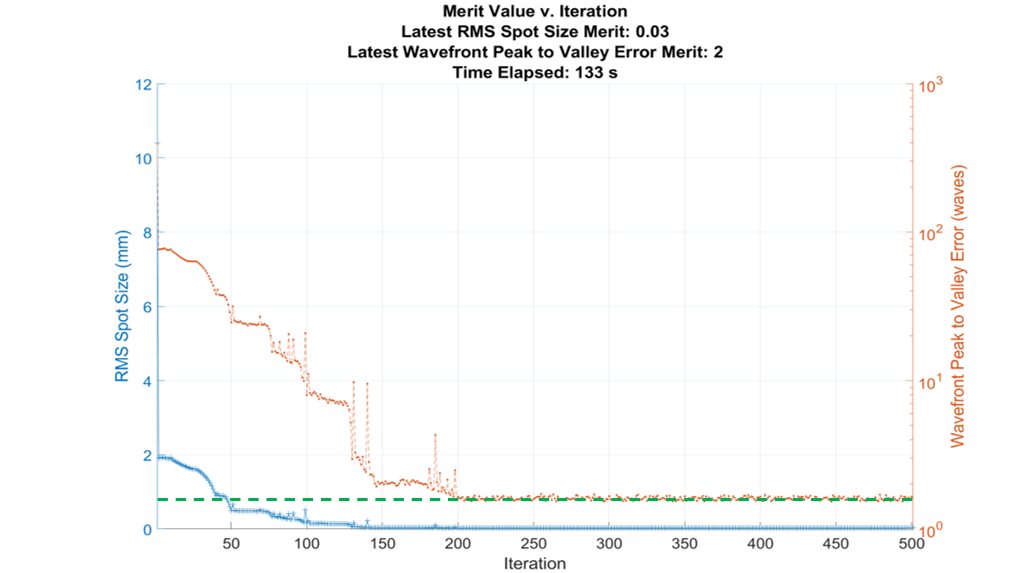}
    \label{fig:enter-label}
    \caption{Plot showing the convergence of RMS spot size radius (mm) and wavefront peak-to-valley merit value (waves) at 650 nm wavelength. The final values after 500 iterations of SPGD are 0.03 mm (average RMS spot size radius) and 2 waves (wavefront peak-to-valley error). The green dashed horizontal line shows the requirement SPGD must reach for further alignment.}
\end{figure}

\clearpage

\section{Phase II Misalignment and Correction}

 For Phase II of the SPGD algorithm, we will consider misalignment for both M1 and M2 but only use the five degrees of freedom on M2's position to correct the system after misaligning both M1 and M2. This phase of developing the correction algorithm will help inform us of the limits for how much M1 can be misaligned and still be corrected with M2 and required M2 full stroke of motion. This phase of development for the algorithm is to finer refine the algorithm by assessing symmetry and convergence in field points when aligning M2. When finding optimal Decenter Z (or the separation between M1 and M2), we will still be using the average RMS spot size radius across all field points since defocus is not a field dependent aberration. Decenter X/Y will be corrected using a polynomial fitting technique. The polynomial will be created by moving the M2 hexapod by some constant step and measuring the difference between the center field and edge right (for decenter X) or far up (for decenter Y) field. By moving M2 in some constant direction, we can move through some minimum difference. The relation between the movement of the M2 position and the field point differences is then fitted to a polynomial and the calculated optimal position of M2 provides almost similar misalignment mount for the SPGD alignment. 

 As shown in Figure 3, we are able to successfully converge down to our 0.05 mm requirement (2 lambda PV WFE) within 200 iterations. As shown in the first iteration of the algorithm, the model-based approach brings the quality metric down to $<1$ mm. Regardless of the amount misalignment, the model-based calculation can provide a $<1$ mm starting point for the following calculation. The consistent level of misalignment would guarantee stable and quick convergence in stochastic searching algorithm. The iterations following the model-based approach continue convergence via the algorithm used in SPGD Phase I. The initial misalignment corresponding to Figure 3 is shown in Table 2. With these results for SPGD Phase II, we have successfully shown that given some misalignment for M1 and M2, we can align the TMA telescope system by adjusting the position of M2 which will be informed by the model-based and randomized approach shown in Figure 4. Table 3 shows the final absolute position of M2 by the end of the convergence plot after reaching our target 0.05 mm average spot size radius.
 
 \begin{table}[hbt!]
    \setlength{\tabcolsep}{10pt} 
    \renewcommand{\arraystretch}{1.5} 
    \centering
    \begin{tabular}{|c|c|c|} \hline 
         \textbf{Degree of Freedom (DoF)}& \textbf{M2 Misalignment (units)} & \textbf{M1 Misalignment (units)}\\ \hline 
         Decenter X & 3 (mm) & 5 (mm) \\ \hline 
         Decenter Y & 3 (mm) & 5 (mm) \\ \hline 
         Separation between M1 and M2 & 3 (mm) & 5 (mm) \\ \hline 
         Tilt X & 0.05 (deg.) & 0.01 (deg.) \\ \hline 
         Tilt Y & -0.05 (deg.) & -0.01 (deg.) \\ \hline 
    \end{tabular}
    \caption{Initial misalignment values applied to M1 and M2 before the SPGD optimization algorithm. This misalignment corresponds to the initial misalignment of the convergence plot shown in Figure 3.}
    \label{tab:my_label}
\end{table}

\begin{table}[hbt!]
    \setlength{\tabcolsep}{10pt} 
    \renewcommand{\arraystretch}{1.5} 
    \centering
    \begin{tabular}{|c|c|c|} \hline 
         \textbf{Degree of Freedom (DoF)}& \textbf{Final M2 Position (units)} \\ \hline 
         Decenter X & 6 (mm) \\ \hline 
         Decenter Y & 6 (mm) \\ \hline 
         Separation between M1 and M2 & 5 (mm) \\ \hline 
         Tilt X & 0.04 (deg.) \\ \hline 
         Tilt Y & -0.04 (deg.)  \\ \hline 
    \end{tabular}
    \caption{Final absolute position of M2 after the correction performed with the algorithm described in SPGD Phase II.}
    \label{tab:my_label}
\end{table}

\begin{figure}[hbt!]
\begin{subfigure}{0.9\textwidth}
    \centering
    \includegraphics[width=0.9\linewidth]{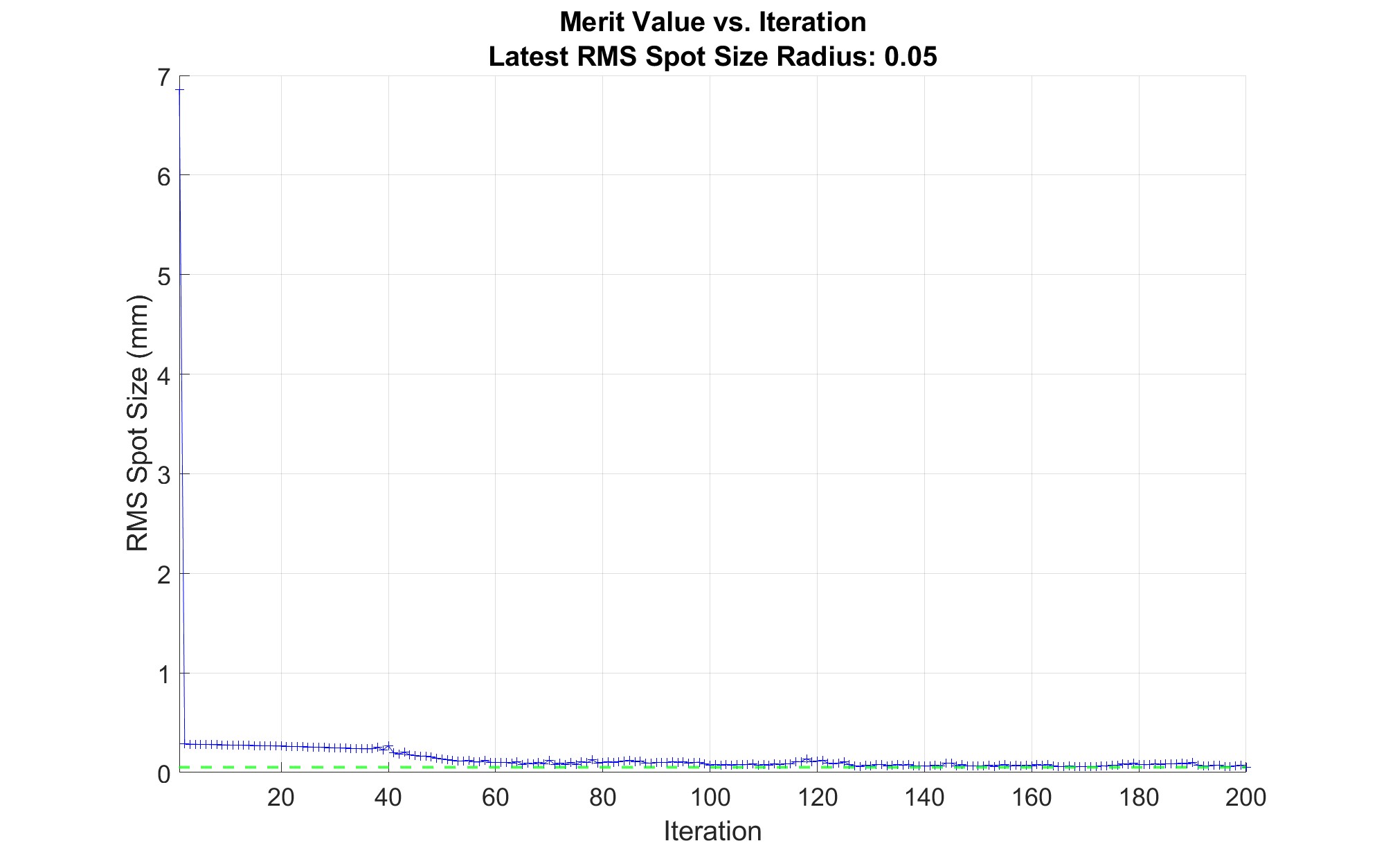}
    \label{fig:enter-label}
    \caption{}
\end{subfigure}
\begin{subfigure}{0.9\textwidth}
    \centering
    \includegraphics[width=0.9\linewidth]{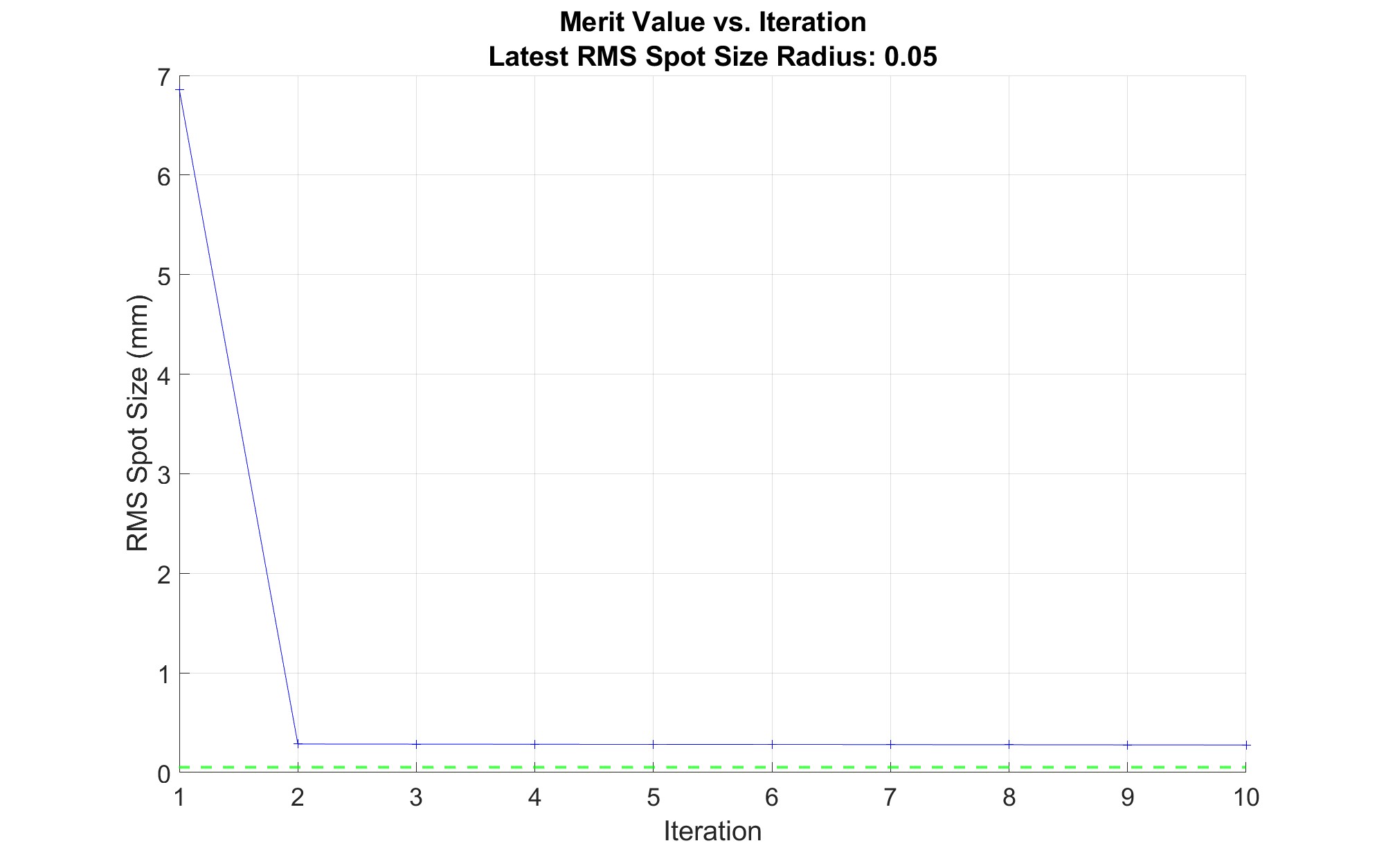}
    \caption{}
\end{subfigure}
    \caption{Plots showing the convergence of RMS spot size radius (mm). The final values after ~200 iterations of SPGD are 0.05 mm (average RMS spot size radius). Plot (a) shows the overall convergence plot over the 200 iterations. Plot (b) shows the first 10 iterations of the algorithm. The green dashed line shows the requirement that SPGD Phase II needs to reach in average spot size radius in order for the next phase of alignment to occur.}
\end{figure}

\clearpage

\begin{sidewaysfigure}[hbt!]
    \centering
    \includegraphics[width=1\linewidth]{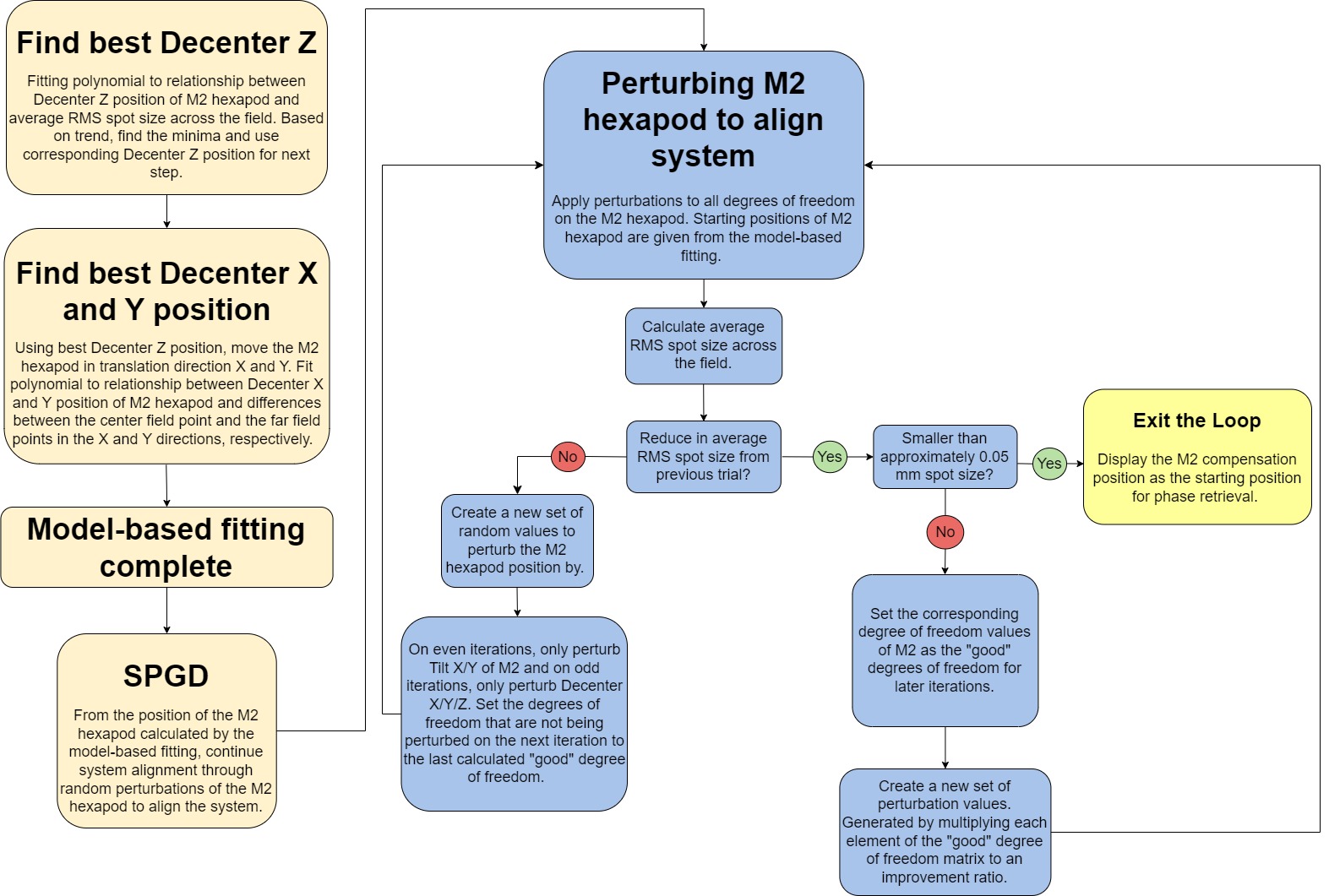}
    \label{fig:enter-label}
    \caption{Flowchart showing the loop of the SPGD process. The first half of the flowchart shows the model-based fitting approach and then followed by the random perturbation process. The entirety of the flow chart is the algorithm that is used when performing SPGD Phase II, including both the model-based and random perturbations. In Phase I, only the random perturbations applied to M2 are used in the SPGD algorithm.}
\end{sidewaysfigure}
 
\clearpage

\section{Conclusions}

This study demonstrates a viable method of coarse alignment for a three mirror anastigmat telescope in the case of static misalignment error. The two phases of development shown above demonstrate the capability of SPGD to be able to correct M1 and M2 misalignments after the telescope reaches its final destination and before fine wavefront sensing and capturing science data. There is further work to be done when considering the mechanical capabilities and limits of M2 during this process. Future work will include implementing the mechanical limits of the M2 range of motion and resolution and using M1 to align balance out the stroke range with M2.  The possibilities of misalignment cases should also be further considered using results from a Monte Carlo simulation to test the robustness of the algorithm for the full range of misalignment cases. We will continue to optimize this algorithm with different misalignment cases and implement the dynamic range of M2 to reflect how this algorithm can be a useful tool when considering an alignment procedure for a TMA telescope design. 

\section*{ACKNOWLEDGEMENTS}

Portions of this research were supported by funding from the Technology Research Initiative Fund (TRIF) of the Arizona Board of Regents and by generous philanthropic donations to the Steward Observatory of the College of Science at the University of Arizona.

\nocite{*}
\bibliography{report.bib} 

\begin{thebibliography}{1}

\bibitem{martin_fabrication_1997}
Martin, H.~M., Burge, J.~H., Ketelsen, D.~A., and West, S.~C., ``Fabrication of the 6.5-m primary mirror for the {{Multiple Mirror Telescope Conversion}},'' in [{\em Optical {{Telescopes}} of {{Today}} and {{Tomorrow}}}{\nolinebreak\hspace{0.1em}]},  Ardeberg, A.~L., ed.,  399--404 (Mar. 1997).

\bibitem{kim2023compact}
Kim, D., Choi, H., and Douglas, E., ``Compact three mirror anastigmat space telescope design using 6.5 m monolithic primary mirror,'' in [{\em Astronomical Optics: Design, Manufacture, and Test of Space and Ground Systems IV}{\nolinebreak\hspace{0.1em}]},   {\bf 12677},  154--159, SPIE (2023).

\bibitem{vorontsov1998stochastic}
Vorontsov, M.~A. and Sivokon, V.~P., ``Stochastic parallel-gradient-descent technique for high-resolution wave-front phase-distortion correction,'' {\em JOSA A}~{\bf 15}(10),  2745--2758 (1998).

\bibitem{blomquist2023analysis}
Blomquist, S., Martin, H., Kang, H., Whitsitt, R., Derby, K., Choi, H., Douglas, E.~S., and Kim, D., ``Analysis of active optics correction for a large honeycomb mirror,'' in [{\em Astronomical Optics: Design, Manufacture, and Test of Space and Ground Systems IV}{\nolinebreak\hspace{0.1em}]},   {\bf 12677},  172--179, SPIE (2023).

\end{thebibliography}
\bibliographystyle{spiebib} 

\end{document}